\documentclass[aps]{revtex4}
\usepackage{epsfig,graphics,amssymb,amsmath,subeqnarray,amsthm}

\def\p{\partial}
\def\d{{\rm d}}
\def\dg{\dot{\gamma}}
\def\e{\epsilon}
\def\g{\frac{\d p}{\d x}}

\def\sb{\bar{\sigma}}
\def\sign{\frac{|\sb(x)|}{\sb(x)}}

\begin{document}

\title{\textbf{Tuning gastropod locomotion: Modeling the influence of \\ 
mucus rheology on the cost of crawling
}}
\author{Eric Lauga\footnote{Email: lauga@mit.edu}}
\affiliation{Department of Mathematics,
Massachusetts Institute of Technology,
77 Massachusetts Avenue,
Cambridge, MA 02139.}
\author{A. E. Hosoi}
\affiliation{Hatsopoulos Microfluids Laboratory, 
Department of Mechanical Engineering, 
Massachusetts Institute of Technology,
77 Massachusetts Avenue,
Cambridge, MA 02139.}
\date{\today}
\begin{abstract}

Common gastropods such as snails crawl on a solid substrate by propagating muscular waves of shear stress on a viscoelastic mucus. Producing the mucus accounts for the largest component in the gastropod's energy budget, more than twenty times the amount of mechanical work used in crawling. Using a simple mechanical model, we show that the shear-thinning properties of the mucus favor a decrease in the amount of mucus necessary for crawling, thereby decreasing the overall energetic cost of locomotion. 

\end{abstract}
\maketitle

\section{Introduction}

Common gastropods crawl on land by propagating waves of shear stress, driven by alternating regions of muscular contraction and expansion, on top of a thin film of viscoelastic mucus released from their foot \cite{morton64,Miller74a,Miller74b,jones75}. The only interactions between the organism and the substrate are through this thin mucus layer, hence locomotion is made possible solely via the fluid dynamics within this thin film.  The nonlinear rheological properties of the mucus are responsible for the remarkable ability of these mollusks to walk on solid ground without detaching the foot from the substrate \cite{denny80a,denny80c}. Mucus production accounts for about a third of the total energy budget of the animal, and is  an order of magnitude larger than the mechanical work required for locomotion, making slug crawling the most energetically expensive mode of locomotion known among vertebrates and invertebrates \cite{denny80b,davies90,kideys91,davies98}. In this paper, we  address the relationship between the cost of locomotion and the mechanical properties of the mucus.

Locomotion strategies employed by gastropods have been of interest to the scientific community for more than  a hundred years, starting with the works of  Simroth \cite{simroth}, Vl\`es \cite{Vle:07} and Parker \cite{parker11}; the reader is referred to Refs. \cite{morton64,gray68,Miller74a,Miller74b,jones75,trueman75,morton79} for reviews. 
Gastropods possess a single flexible foot enhanced with arrays of cilia. In many cases, the  motion of the cilia is responsible for movement of the mucus underneath the gastropod and the resulting motion of the animal. However, many gastropods, including all snails that crawl on hard substrates, use a second mechanism to crawl, namely the actuation of muscular pedal waves. Using a series of  foot muscles, gastropods are able to stress the thin layer of mucus (which is typically on the order of tens of microns thick) with alternating regions of muscular contraction and expansion. These regions oscillate in time, leading to traveling waves  that shear the mucus, resulting in translation of the animal with crawling velocities typically between 1 mm/s and 1 cm/s.  The muscular  waves are said to be direct if they propagate in the same direction as the direction of locomotion, and retrograde otherwise. Direct waves are waves of contraction, retrograde waves are waves of extension.  Similarly, waves are classified as monotaxic when a single wave spans the foot of the animal and ditaxic if if two alternating waves span the foot.   In the three large classes of gastropods, pulmonates are (mostly) terrestrial crawlers and use direct monotaxic waves. Prosobranchs represent the largest class of gastropods and are the most varied in structure; most of them are marine gastropods using ditaxic retrograde waves. Finally, opisthobranchs do not crawl, but burrow or swim \cite{farmer70}.

Lissman was the first to study in detail the kinematics and dynamics of gastropod motion \cite{Lis:45,Lis:46}. He analyzed the locomotion of three pulmonates, {\it Helix aspera} (monotaxic direct), {\it Haliotis tuberculata} and {\it Pomatias elegans} (both ditaxic direct).  Jones and Trueman \cite{jones70,trueman84} were the first to study retrograde wave locomotion, on {\it Patella vulgata} (ditaxic retrograte).  However, it was not until 1980 with the work of Denny and co-authors that a complete mechanical picture for gastropod locomotion emerged.
The key feature of gastropod locomotion lies in the mechanical properties of the mucus layer. This was first discovered for the Pumonate slug  {\it Ariolimax columbianus} \cite{denny80a,denny80c}, but is generally valid for all gastropods \cite{skingsley00,smith02} (see also \cite{denny81,denny83,denny84,denny89}). The mucus is composed of more than 95\% water,  some dissolved salts, and the remaining 3-4\% is a high molecular weight glycoprotein (mucin) which,  in solution, forms a cross-linked gel network. This network is responsible for the elastic component of the mucus. At small  shear strains, the mucus behaves like an elastic solid, but at large shear strains (on the order of 5), the mucus yields and behaves like a viscous liquid, with the shear viscosity dropping by about three orders of magnitude to a value  of approximately twenty times that of water. After yielding, the mucus locally heals back into a gel network on a time scale of less than 0.1 s
\footnote{These mechanical  properties are very similar to that of other forms of animal secretions (respiratory, gastric and cervical mucuses).}. Underneath a crawling snail,  shear strains vary in time between 0, in the so-called ``interwave" regions in which the foot does not move relative to the ground, and approximately 50 in the regions of largest relative displacement. It is this combination of  yield-stress properties with judicious periodic  shearing of the mucus that allows parts of the foot to adhere to the ground while other portions are displaced, leading to an overall translation of the organism.

The use of a viscoelastic mucus for locomotion has a second interesting consequence. Measurements of the net increase in oxygen consumption due to locomotion for {\it Ariolimax columbianus} showed that adhesive crawling is the most expensive mode of locomotion known \cite{denny80b},  twelve times as costly as running for example. Perhaps surprisingly, the largest component of this energy budget does not arise from the mechanical work of the muscles, but is due to the chemical cost of mucus production (that is, the chemical cost of creating  the glycoprotein based on its  sequence of amino-acids). For {\it Ariolimax columbianus}, mucus production is about twenty times more costly than the mechanical work of locomotion \cite{denny80b}.  Since Denny's seminal work, the cost of locomotion has been analyzed for other gastropods, including {\it Patella vulgata} \cite{davies90}, {\it Buccimum undatum} \cite{kideys91} and {\it Haliotis kamtschatkana} \cite{donovan97} and similar results have been obtained (see also  \cite{davies98}) .

In this paper, using a simple model for the nonlinear rheology of the mucus, we show that its shear-thinning properties allow the gastropod to decrease the aforementioned energetic cost of locomotion by minimizing the amount of fluid dispensed.
The paper is organized as follows. We introduce our model for the nonlinear rheological properties of the mucus in \S\ref{model},  set up the geometry and notation for the particular mathematical problem to be solved in \S\ref{setup} and present its  asymptotic solution in \S\ref{asymptotic}. We apply it to the typical stress distribution underneath a crawling slug in \S\ref{punchline} and show that choosing the mucus to be shear-thinning allows to decrease the amount of required fluid.

\section{Model}
\label{model}

\begin{figure}[t]
\centering
\includegraphics[width=.4\textwidth]{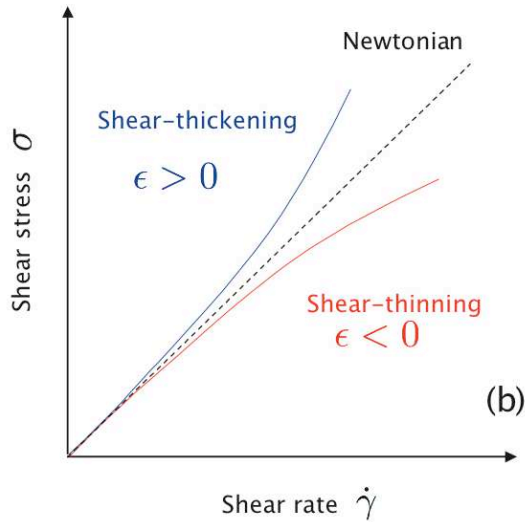}
\caption{
Schematic representation of the model used to represent a slightly non-Newtonian mucus in this paper. The straight line represent a Newtonian mucus; when the stress-shear rate curve is above the Newtonian one ($\epsilon > 0$) the mucus is shear-thickening, and when it is below ($\epsilon < 0$) the mucus is shear-thinning.}
\label{data}
\end{figure}

Adhesive locomotion, the crawling strategy employed by gastropods, has therefore two distinctive features: (1) The nonlinear rheology of the mucus, solid-like at low stresses and liquid-like for larger stresses, and (2) the high cost of locomotion due to the internal chemical production of the proteins which compose the mucus. In this paper, we  address the possible link between these two features. Generically, a gastropod should be able to crawl on any kind of non-Newtonian mucus, whether it is shear-thinning (as a dilute polymer solution), shear-thickening (as cornstarch in water), or any other rheology  characterized by a nonlinear relationship between stress and shear rates. 

The mucus of real snails possesses a yield stress, as demonstrated 
by the experimental work of Denny  \cite{denny80a,denny80c}, and is therefore abruptly shear-thinning with shear viscosities dropping by more than three orders of magnitude around the yield point  \cite{Randy}.
Here we address the question: Could it be that this particular rheology has been tuned to allow locomotion at the lowest energetic cost to the animal? To understand the evolutionary process that could lead to such a mucus, consider an organism mechanically similar to a snail that secretes a simple lubricating fluid (e.g.~water).  This ÒmucusÓ becomes more complex as its chemical composition evolves.  Hence, as its rheology is perturbed away from Newtonian, it is necessary to elucidate which changes to the rheology are beneficial to the organism. This is the problem we propose to address, perhaps shedding some light on the importance of material properties in an evolutionary context.

Guided by this evolutionary picture, we make the following modeling assumption: Let us consider the  mucus of this primitive snail-like organism as slightly non-Newtonian, as illustrated in Fig.~\ref{data}, with a quadratic relationship between shear rate, $\dot{\gamma}$ ($=\p u / \p y$), and shear stress, $\sigma$, given by
\begin{equation}
\dg = \frac{\sigma}{\mu} \left(1-\e \frac{|\sigma|}{\sigma_*} \right) \cdot
\label{shear}
\end{equation}
As a first approach to the problem, we model therefore  the mucus as purely viscous (generalized Newtonian) and neglect its elastic properties. Our empirical quadratic model (Eq.~\ref{shear}), a Cross model of order one, is equivalent to a Taylor expansion at small stress of the inverse of the shear viscosity, $\eta$ ($= \sigma /  \dot{\gamma}$), as a function of the absolute value of the shear stress, 
and is arguably the simplest model that captures the effects  of the nonlinear rheology \footnote{The absolute value in the expansion has been included to preserve the physical symmetries in the problem.  To demonstrate that our results do not rely on the singularity arising from this absolute value, we present an alternate analytic expansion in Appendix \ref{analytic}.}. 
Here $\sigma_* > 0$ is a typical shear stress, $\mu$ the viscosity in the limit of small stress and $\epsilon$ a small parameter, $|\epsilon| \ll1$, whose absolute value quantifies the departure from Newtonian behavior and whose sign determines the rheological properties ($\epsilon < 0$ corresponds to shear-thinning and  $\epsilon >0$ to shear-thickening) and will be chosen to minimize the cost of locomotion. 

As we have seen before, this cost is related to the cost of producing the mucus which differs from the traditional measure of energy expenditure associated with mechanical work. In real gastropods, the mucus is dispensed from membrane-bound vesicles at various location along foot \cite{deyrupolsen83}. Let us denote by $e$ the cost of fabricating the mucus per unit mass, $\rho$ the mucus density and $Q_s$ the flow rate of the mucus flowing underneath the snail and measured in a frame moving with the snail. Then the cost of locomotion, ${\cal W}$, over a time interval $\Delta t$, is given by 
\begin{equation}
{\cal W}=e \rho \Delta t Q_s,
\end{equation}
and minimizing this cost is therefore equivalent to minimizing the required amount of mucus (per time), $Q_s$.  

In the reminder of this paper, we solve the equation of motion for the crawling snail perturbatively in $\epsilon$ and find which material properties minimize the flow rate of the mucus, thereby minimizing the overall cost of locomotion.

\subsection{Equations of Motion}
\label{setup}

\begin{figure}[t]
\centering
\includegraphics[width=.95\textwidth]{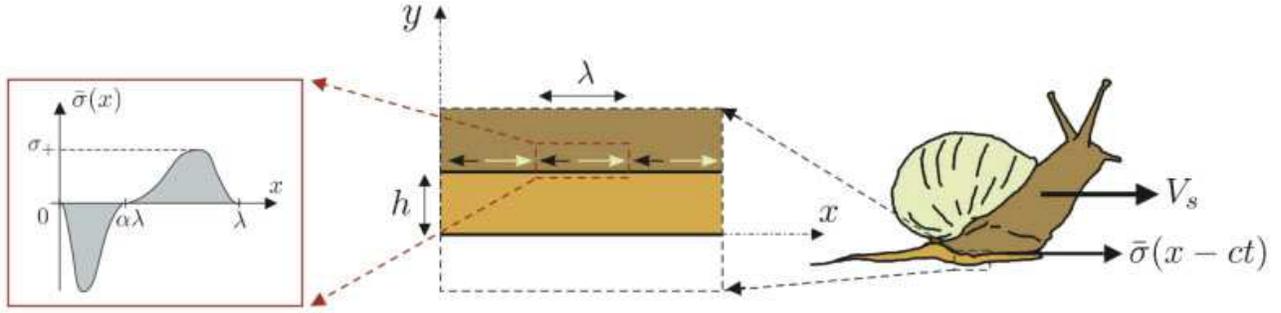}
\caption{Notation for the model snail. Details are given in \S\ref{setup}.}
\label{mainfigure}
\end{figure}

Consider a snail that crawls in the $x$-direction on top of a thin liquid film of mucus of constant thickness $h$ (see Fig.~\ref{mainfigure}).  At $y=h$, the gastropod exerts on the mucus, in a frame moving with the snail, a known traveling wave of  shear stress, $\sigma_{xy}(x,h,t)=\sb(x-ct)$, where $c$ is the wave velocity as dictated by the muscular contractions. At $y=0$, the mucus adheres to the substrate and satisfies a no-slip boundary condition. We denote the unknown instantaneous crawling speed of the center of mass of the gastropod by $V_s{\bf e}_x$, where ${\bf e}_x$ is the unit vector in the $x$-direction.
In the reference frame moving with the shear wave, the mechanical problem  is time-independent. In this frame, equilibrium for the incompressible mucus at low Reynolds numbers can be written as
\begin{equation}
\nabla p = \nabla\cdot \boldsymbol{\sigma},\quad\nabla\cdot {\bf u}=0
\end{equation}
where $p$ denotes the pressure field, ${\bf u}$ the velocity field  and $\boldsymbol{\sigma}$ the deviatoric stress tensor, 
subject to the boundary conditions
\begin{equation}
\sigma_{xy} (x,h)=\sb(x) \quad \mbox{and} \quad {\bf u}(x,0)=-c-V_s,
\end{equation}
where ${\bf u}$ denotes the velocity field. The equations above are to be solved subject to three constraints: (1) The organism is force-free; (2) by periodicity, there is no net build-up of pressure over one wavelength of muscular action within the mucus layer; (3) in the frame moving at speed $V_s{\bf e}_x$, the instantaneous velocity of the center of mass of the gastropod is zero (by definition of $V_s$) i.e.
\begin{equation}\label{constraints}
\int_0^\lambda \sb(x) \, \d x = 0, \quad  \int_0^\lambda \frac{\p p}{\p x}(x,y) \, \d x=0,\quad \mbox{and} \quad \int_0^\lambda [u(x,h)+c] \,\d x= 0
\end{equation}
where $\lambda$ is the wavelength of the periodic muscular shearing of the mucus.
We further simplify the problem by noting that $h\ll \lambda$, and therefore we can apply a lubrication approximation. Consequently, the only relevant shear stress is $\sigma_{xy}$, which we denote as 
$\sigma$ for simplicity, and the only relevant velocity component is $u={\bf u}\cdot {\bf e}_x$. The equations for steady state crawling then become 
\begin{equation}
\frac{\p p}{\p x}  =  \frac{\p \sigma }{\p y}, \ \ \ \ \ \ \ 
\frac{\p p}{\p y}  =  0.
\end{equation}
Using the stress boundary condition, these equations can be integrated once to obtain
\begin{equation}
\label{stress}
\sigma(x,y)=\g (y-h) + \sb(x)\cdot
\end{equation}
The shear stress in the mucus, given by Eq.~\eqref{stress}, is therefore a linear function of $y$.  We will denote  the coordinate where the shear stress  changes sign by $y^*(x)$, that is 
\begin{equation}\label{ystar}
\sigma(x,y)=\g (y-y^*).
\end{equation}
Note that above $y^*$, the stress has the same sign as $\sb$, and below the opposite sign. As will be seen below, the leading-order term for $y^*$ is constant and equal to $h/3$, so that the stress does indeed change sign within the mucus.
Finally, the first two constraints given in Eq.~\eqref{constraints} lead to 
\begin{equation}
\int_0^\lambda \g\, y^* \d x= 0.
\end{equation}

\subsection{Asymptotic Solution}
\label{asymptotic}
It is now possible to solve for the dynamics of the crawling motion.
From Eq.~\eqref{shear} and \eqref{ystar}, the equilibrium equations to be solved become
\begin{equation}\label{tosolve}
\frac{\p u }{\p y }   =  \frac{1}{\mu} \g (y-y^*) +  \epsilon \frac{\mathrm{sgn}\left(y^*-y\right)}{\mu \sigma_*} \left(\g\right)^2 \sign (y-y^*)^2.
\end{equation}
Using the boundary condition $u(x,0)=-c-V_s$, and the fact that the velocity profile must be continuous at $y^*$, Eq.~\eqref{tosolve} can be integrated to give
\begin{subeqnarray}
u(x,y) & = & \frac{g}{2\mu} y(y-2y^*) + \epsilon \frac{g^2 S}{3\mu \sigma_*}  [(y-y^*)^3+y^{*3}] - c-V_s, \quad y \leq y^*\\
u(x,y) & = &  \frac{g}{2\mu} y(y-2y^*) - \epsilon \frac{g^2 S}{3\mu \sigma_*}  [(y-y^*)^3-y^{*3}] - c-V_s, \quad y \geq y^*
\end{subeqnarray}
where we have defined $\g \equiv g(x)$ and $\sign \equiv S(x)$ in order to simplify the notation.
The mucus velocity at $y=h$ relative to the moving gastropod is therefore given by
\begin{equation}
u_s(x) \equiv u(x,h)+ c = \frac{hg}{2\mu} (h-2y^*) - \epsilon \frac{g^2 S}{3\mu \sigma_*}  [(h-y^*)^3-y^{*3}] -V_s,
\end{equation}
and the flow rate in the frame moving with the gastropod is given by
\begin{equation}\label{flowrate}
Q_s=\int_0^h [u(x,y) + c] \,   \d y = \frac{gh^2}{2\mu}\left(\frac{h}{3}-y^*\right) - V_sh - \frac{\e g^2 S}{12\mu \sigma_*} [y^{*4} + (h-y^*)^4-4y^{*3}h].
\end{equation}
Finally the average rate of energy dissipated through viscosity (which is approximately equal to the rate of mechanical work done by the gastropod to crawl) is given by 
\begin{equation}\label{work}
{\cal E} = \int_0^\lambda \int_0^h  \sigma \dot{\gamma}\,\d y\, \d x.
\end{equation}
The solution can now be expanded in powers of $\e$ and solved order-by-order.  For details see Appendix \ref{order-by-order}.
Recall that our goal is to find the influence of the sign of $\epsilon$ on (1) the chemical work required to produce the mucus, which is proportional to $Q_s$ as defined in Eq.~\eqref{flowrate}, and (2) the mechanical rate work of crawling, as defined in Eq.~\eqref{work}.

At lowest order, the flow is Newtonian.
The constraints are written at this order as
\begin{equation}
\int_0^\lambda g_0\,  \d x= 0 \quad \mbox{and} \quad \int_0^\lambda g_0 y_0 \, \d x= 0,\quad \int_0^\lambda u_0(x,h)\,\d x = 0 .
\end{equation}
Averaging the zeroth order velocity field (Eq.~\ref{u0}) over one wavelength and applying the constraints leads to a lowest order crawling velocity $V_0=0$.
Thus, as expected,  the organism cannot propel itself on a Newtonian fluid. 
Averaging the volume flux (Eq.~\ref{Q0}) over one wave length, gives $Q_0 =0$, and therefore $y_0 = h/3$. Using this result in Eq.~\eqref{ystar} we find the pressure gradient and the energy associated with mechanical work are given by
\begin{equation}
g_0(x) = \frac{3\sb(x)}{2 h }
\quad \mbox{and} \quad
{\cal E}_0  =   \frac{h \lambda }{4\mu}\langle \sb^2 \rangle >0,
\end{equation}
respectively, where we have denoted
$\langle f \rangle \equiv \frac{1}{\lambda}\int_0^\lambda f(x) \d x = \int_0^1 f(\lambda u)\d u.$

At  order $\e$ we have to enforce the constraints
\begin{equation}
\int_0^\lambda g_1 \d x= 0,\quad \int_0^\lambda [g_0  y_1 + g_1 y_0] \d x= 0,\quad \int_0^\lambda u_1(x,h) \d x = 0 .
\end{equation}
Averaging the velocity at this order (Eq.~\ref{u1}) leads to a non-zero first order velocity
\begin{equation}
V_1=-\frac{7}{36}\frac{h}{\mu \sigma_*} \langle \sb | \sb| \rangle.
\label{V1}
\end{equation}
So in general, the organism can crawl at order $\e$. The motion is retrograde if $\epsilon \langle \sb | \sb| \rangle > 0$, and direct otherwise. 
Averaging $Q$ at order $\e$ (Eq.~\ref{Q1}) then gives the first order mucus flow rate 
\begin{equation}
Q_1=\frac{79}{432}\frac{h^2}{\mu \sigma_*} \langle \sb | \sb| \rangle.
\end{equation}
Note that $V_1Q_1 < 0$ as is expected.
Substituting this result back into the first order flux equation (Eq.~\ref{Q1}) (recall $Q_s$ is a constant independent of $x$) and using both the lowest order solution for $y_0$ and Eq~\eqref{V1} leads to
$g_0 y_1 =  5( \langle \sb | \sb| \rangle- \sb | \sb|)/(216 \sigma_*)$.
Expanding Eq.~\eqref{ystar} at order  $\e$ gives
$g_1= 3 g_0 y_1 / 2h$, leading to the first order term in the pressure gradient
\begin{equation}
g_1 = \frac{5}{144}\frac{1}{h \sigma_*} ( \langle \sb | \sb| \rangle- \sb | \sb|).
\end{equation}
Finally, we find the first order correction to the mechanical energy
\begin{equation}
{\cal E}_1  =  -\frac{17}{96}\frac{h \lambda }{\mu\sigma_*}\langle \sb^2|\sb| \rangle < 0.
\end{equation}
Consequently, for a known traveling wave of shear stress  applied by the gastropod on the mucus, the least amount of {\em mechanical} work is done when  $\e >0$, that is when the mucus is shear-thickening.

Since we find that $Q_0=0$, we have to calculate the second order correction to the flux in order to quantify the influence of the sign of $\epsilon$ on the {\em chemical} cost of locomotion. At this order we have to enforce the constraints
\begin{equation}
\int_0^\lambda g_2 \d x= 0,\quad \int_0^\lambda[ g_0 y_2 + g_2 y_0 + g_1 y_1] \d x= 0,\quad \int_0^\lambda u_2(x,h) \d x = 0 .
\end{equation}
Again, averaging the velocity at order $\e^2$ (Eq.~\ref{u2}) we find the second order correction to the crawling velocity
\begin{equation}
V_2=\frac{5}{486}\frac{h}{\mu \sigma_*^2} [\langle | \sb| \rangle \langle \sb | \sb| \rangle - \langle \sb^3 \rangle],
\end{equation}
and averaging $Q$ at order $\e^2$ (Eq.~\ref{Q2}) gives  the second order correction to the flux
\begin{equation}
Q_2=\frac{185}{46656}\frac{h^2}{\mu \sigma_*^2}[ \langle \sb^3 \rangle - \langle | \sb| \rangle \langle \sb | \sb| \rangle].
\end{equation}

\section{Relevance to gastropod crawling: Impact of rheology on locomotion cost}
\label{punchline}
We are now equipped to address the main question raised in this paper, as presented in \S\ref{model} namely, have the material properties of pedal mucus been tuned to optimized crawling efficiencies. In the frame moving with the gastropod, the flow rate is given by $Q_s=\e Q_1 + \e^2 Q_2$, with errors of  order $\e^3$. This flow rate is, as discussed above, directly related to the cost of locomotion associated with the chemical production of the mucus. A change of sign for $\e$ modifies the sign of the first order term, but does not change the second order term, and therefore the absolute value of the flow rate is minimized for a particular choice of the sign of  $\e$.  Note that we consider the absolute value of the flow rate to allow for both direct and retrograde waves; i.e.~the direction of motion relative to the direction of the traveling wave is irrelevant.  The absolute value of the flow rate normalized by the first order term is given by
\begin{equation}
\label{asbQ}
\frac{|Q_s|}{| \e Q_1|}= 1 +\frac{\Delta Q}{ Q},
\end{equation}
with
\begin{equation}\label{dQ}
\quad  \frac{\Delta Q}{ Q} = \e \frac{Q_2}{Q_1}
=\frac{185 \epsilon}{8532\sigma_*} \left(\frac{\langle \sb^3 \rangle - \langle | \sb| \rangle \langle \sb | \sb| \rangle}{\langle \sb | \sb| \rangle}\right),
\end{equation}
which we estimate below.

Let us consider the distribution of shear stress underneath a crawling gastropod, $\sb(x)$, and let us zoom in on the details of one wavelength $\lambda$ of the  foot, as illustrated on the left in Fig.~\ref{mainfigure}. We know that the stress, which averages to zero, will be negative in some region of the shear wave and positive in the rest. We can safely assume that it is a smooth function of space, and we write for convenience 
\begin{subeqnarray}
\label{def}
\sb(x)  & = & -\sigma_- g(x),\quad {\rm for }\quad  x\in [0,\alpha \lambda],\\
\sb(x) & = & \sigma_+ f(x),\quad  {\rm for }\quad x\in [\alpha \lambda ,\lambda],
\end{subeqnarray}
where $f$ and $g$ are positive functions with maximum value 1, $\alpha \in [0,1]$ represents the fraction of the wavelength which has negative stress, and $\sigma_-$ and $\sigma_+$ are the maximum absolute values of the regions of negative and positive stresses respectively. 
As indicated in Fig.~\ref{mainfigure}, we have defined the wavelength such that $\sb(0)=\sb(\lambda)=0$ and, given Eq.~\eqref{def}, the stress is also equal to zero at $x=\alpha\lambda$. Physically,  the stress curve in Fig.~\ref{mainfigure} can be thought of as the profile of the muscular strains in the foot, and therefore the area where the stress $\sb$ goes from negative to positive (around $x=\alpha \lambda$) is a region of muscular extension  and that where  $\sb$ goes from positive to negative  (around $x=0$ and $\lambda$) is an area of muscular contraction.  

The profile of shear stress exerted by the gastropod on the mucus is a result of the action of its foot muscles.  The most detailed studies of foot muscles in real snails are due to Jones and coauthors, for  {\it Patella vulgata}  \cite{jones70} and {\it Agriolimax reticulatus} \cite{jones73}. They showed that the distribution of muscle fibers in gastropod varies widely between species, as confirmed by subsequent studies of  {\it Neritina reclivata} and {\it Thais rustica}  \cite{gainey76},  {\it Helix aspersa} \cite{buyssens04}, and  {\it Melampus bidentatus} \cite{moffett79}. However, they also showed that (1)  the distribution of muscles responsible for the wave motion are front-back symmetric in each of the the organisms considered (dorso-ventral muscles for {\it Patella vulgata}  leading to retrograde waves, and longitudinal, transverse and oblique muscles for {\it Agriolimax reticulatus}  and direct waves), and (2) that the same sets of muscles were responsible for both extension and contraction of the foot.  Based on these observations, we will  assume in the remainder of this paper that the action of the muscles result in wave profiles in which the regions of muscular extension and those of muscular contraction have similar profiles. In other words, up to scaling factors in both the $x$ and $y$ directions, regions of positive and negative stress as sketched in Fig.~\ref{mainfigure} will be assumed to have similar shapes. We expect this assumption - the simplest assumption that can be made about the stress profile  in Fig.~\ref{mainfigure} - although it is restricted to organisms with similar characteristics to {\it Agriolimax reticulatus}  and {\it Patella vulgata}, to remain true is other types of gastropods.

The mathematical formulation of this assumption is  given by a relationship between $f$ and $g$ as
\begin{equation}
g(x) = f\left(\frac{\alpha-1}{\alpha}x + \lambda \right)\cdot
\end{equation}
For simplicity we introduce a new positive function, $w(u)$, defined for $u\in [0,1]$ as $w(u) = g[\alpha\lambda (1-u)]=f[\lambda(\alpha + (1-\alpha)u)] $. Since the organism as a whole is force free,
\begin{equation}
\int_0^\lambda  \sb (x)\,  \d x = \lambda \alpha \left(\frac{1-\alpha}{\alpha} \sigma_+ -  \sigma_- \right)
I_1 = 0,
\end{equation}
where we denote
\begin{equation}
I_n \equiv \int_0^1 w^n(u)\,\d u > 0.
\end{equation}
Since $I_n$ is strictly positive, we require
\begin{equation}
\sigma_- =  \frac{1-\alpha}{\alpha}\sigma_+ \cdot
\end{equation}
We can now proceed with the calculation of the various averaged quantities along one wavelength of the foot. 
To calculate the effect of the second order correction to the flux from Eq.~\eqref{dQ}, we require expressions for $\langle \sb | \sb| \rangle$, $\langle \sb^3 \rangle$ and $\langle | \sb| \rangle$.
Rewriting these quantities in terms of $\sigma_+$, $\alpha$ and $I_n$, we obtain
\begin{equation}
\langle \sb | \sb| \rangle = -\sigma_-^2\int_0^{\alpha\lambda}g^2(u)\,\d u + \sigma_+^2 \int_{\alpha\lambda}^\lambda f^2(u)\, \d u
 = \frac{(2\alpha-1)(1-\alpha)}{\alpha}\sigma_+^2  I_2,
\end{equation}
\begin{equation}
\langle \sb^3 \rangle =-\sigma_-^3\int_0^{\alpha\lambda}g^3(u)\,\d u + \sigma_+^3 \int_{\alpha\lambda}^\lambda f^3(u)\, \d u
= \frac{(2\alpha-1)(1-\alpha)}{\alpha^2} \sigma_+^3 I_3,
\end{equation}
\begin{equation}
\langle | \sb| \rangle = \sigma_-\int_0^{\alpha\lambda}g(u)\,\d u + \sigma_+ \int_{\alpha\lambda}^\lambda f(u)\, \d u
=2(1-\alpha)\sigma_+ I_1.
\end{equation}
Thus the excess flow rate, as calculated form Eq.~\eqref{dQ}, is given by
\begin{equation}
 \frac{\Delta Q}{ Q} =\epsilon \frac{185}{4266}\left(\frac{\sigma_+}{\sigma_*}\right)\left(\frac{\alpha^2-\alpha + \beta}{\alpha}\right)I_1
\label{DeltaQ}
\end{equation}
where we have defined $\beta \equiv {I_3}/{2I_1I_2}$.  
Recall that we are interested in the sign of this term to determine which sign of $\e$ is beneficial to the cost of locomotion. Hence we need  to determine the sign of $\alpha^2-\alpha + \beta$, as all of the other terms on the right hand side of Eq.~\eqref{DeltaQ} are strictly positive. It is straightforward to show that this term is positive if $4\beta > 1$ or equivalently, if $2I_3 > I_1I_2$. This inequality can be demonstrated as follows:
\begin{eqnarray}
2I_3 - I_1I_2 & = & \int_0^1 w^2(x)\left[2w(x)-\int_0^1 w(u) \d u\right]\d x \nonumber \\
&=&  \int_{2w\leq \int_0^1 w} w^2(x)\left[2w(x)-\int_0^1 w(u) \d u\right]\d x
+ \int_{2w\geq \int_0^1 w} w^2(x)\left[2w(x)-\int_0^1 w(u) \d u\right]\d x  \nonumber \\
& \geq & \left(\frac{1}{2} \int_0^1 w(u) \d u\right)^2 \left\{\int_{2w\leq \int_0^1 w} \left[2w(x)-\int_0^1 w(u) \d u\right]\d x
+ \int_{2w\geq \int_0^1 w} \left[2w(x)-\int_0^1 w(u) \d u\right]\d x  \right\} \nonumber \\
& \geq & \frac{1}{4}\left(\int_0^1 w(u) \d u \right)^3 > 0. \nonumber 
\end{eqnarray}
We have therefore shown that $\alpha^2-\alpha + \beta > 0$, which implies that the sign of  $\Delta Q / Q $ is the same as the sign of $ \epsilon $, for all possible functions $w$. Given Eq.~\eqref{asbQ}, this means that the absolute value of the flow rate will be minimized if we choose $\e < 0$, i.e.~a shear-thinning fluid. As chemical production represents the largest energy cost associated with locomotion, this criteria trumps the mechanical work result given above, and a shear-thinning fluid will decrease the overall cost of locomotion for the gastropod.

\section{Discussion}

In this study, we have investigated the possible link between rheological properties of gastropod mucus and the large cost of locomotion experienced by these animals. In theory, any type of nonlinear relationship between shear stress and shear rate should allow the animal to move, hence there are a number of open questions regarding the possible mechanisms and criteria used in selecting particular types of mucus. The approach we take is to use a generalized Newtonian model (Fig.~\ref{data} and Eq.~\ref{shear}) to quantify how to best perturb the mucus away from a Newtonian fluid. Using a series of assumptions which are justified from an experimental standpoint, we have been able to show that a mucus which is shear-thinning allows the gastropod to crawl while using the least amount of fluid. As the energetic cost of locomotion for these animals is largely dominated by that of mucus production, a shear-thinning material allows locomotion at the least expense.  This is consistent with experimental observations as all studies which have quantified the rheology of real gastropod mucus have found the fluid to be strongly shear-thinning.

Real mucus is abruptly shear-thinning, so is tempting to suggest that our simple model can provide some rationale for the biological series of events that transformed  a simple lubricating fluid into an extremely nonlinear material through the amplification of its beneficial nonlinear characteristics. Certainly the lowered cost of locomotion is an added benefit for the animal regardless of whether or not the energetics played a significant role in the evolutionary process.
However, our calculations were made under several simplifying assumptions, the most severe of which is the focus on viscous features while ignoring elastic stresses. Consequently, at best, we can conclude that our calculations suggest an interesting relation between locomotion cost and the mechanical properties of the fluid.  

The second  point of interest is that  if, instead of using what we know to be the real measure of locomotion cost for snails -- the chemical cost associated with mucus production --  we use a more traditional measure of locomotion cost --  the mechanical work of the crawling organism -- we find the opposite result. That is, we find that it is beneficial to use a shear-thickening fluid to decrease  the amount of work necessary to move. This erroneous conclusion  emphasizes the importance of biologically relevant measures when addressing optimization and tuning of biological systems, and it suggests that the results of our calculations have some biological relevance. Note that it also implies that for synthetic crawlers, for which mucus production is not an issue, shear-thickening fluids are energetically advantageous \cite{chan05}.

Our third conclusion concerns the speed of the muscular waves.  We have found that crawling speeds and flow rates are independent of the speed $c$ of the muscular waves, which might at first appear counterintuitive as it allows for locomotion when $c=0$. However, we have assumed implicitly in our calculations that the size of the snail foot (and therefore its wavelength) remains constant. Thus we have assumed that the muscular extensional strains in the foot remain small, as is observed experimentally \cite{Lis:45,Lis:46,denny80c}.  In the frame moving with the snail, the typical stress $\sb$ is sustained by the same sets of muscles during a time $t \sim \lambda / c$, and the typical local velocity of the foot relative to the ground is such that $\sb \sim \mu v / h$ so that $v \sim h \sb / \mu$. The maximum longitudinal displacement of elements on the foot is therefore $\ell \sim v t \sim h \sb \lambda / \mu c$, and therefore the typical muscular strain is of order  $\ell / \lambda   \sim h \sb  / \mu c$. Requiring an upper bound on muscular strains offers therefore a lower bound for the wave speed, removing the specious $c=0$ paradox.

Finally, we  discuss the difference between direct and retrograde wave crawling. As we have seen, the crawling speed is given at leading order, by $V_s=- 7\e h\langle \sb | \sb| \rangle / 36 \mu \sigma_*$ and, in theory, both direct ($V_s >0$) and retrograde ($V_s<0$) motion is possible. On the other hand, if the foot of the snail lifts from the substrate as the wave passes -- even slightly, as in most snails -- the contribution of this lifting motion to locomotion always results in a retrograde force due to peristaltic effects \cite{chan05}. The consequence of this result is that, if the snail uses direct crawling yet lifts its foot, the contribution due to the non-Newtonian rheology of the mucus needs to be large enough to overcome the retrograde contribution due to foot lifting. More precisely, if the foot shape is written as $h(x) = h_0[1+\delta v(x)]$, where $\delta \ll 1$ is its amplitude and $v(x)$ its dimensionless profile, using the results in Ref.~\cite{chan05}, we find that the crawling velocity is given, at leading order in both $\e$ and $\delta$, by
\begin{equation}
V_s= 6c\delta ^2 \left[\int v^2 -\left( \int v \right)^2 \right]-\frac{7}{36}\frac{\e h}{\mu \sigma_*} \langle \sb | \sb \rangle,
\end{equation}
and direct wave motion ($V_s >0$) is possible only if the mucus is sufficiently non-Newtonian, namely
\begin{equation}
\e \langle \sb | \sb| \rangle < 0,\quad 
|\e| > \frac{216}{7} \frac{c\delta^2 \mu \sigma_* }{h |\langle \sb | \sb| \rangle|} \left[\int v^2 -\left( \int v \right)^2 \right] \cdot 
\label{last}
\end{equation}
In the case where direct motion is preferable from a muscular standpoint, the result of Eq.~\eqref{last} suggest an additional incentive for the mucus to depart from Newtonian.

\appendix
\section{Asymptotic results to order $\e^2$}
\label{order-by-order}
The solution to Eq.~\eqref{tosolve} is expanded in powers of $\e$ as follows:
\begin{eqnarray}
 \left\{
\begin{array}{c}
 u_s(x) \\
 g(x) \\
 y^*(x) \\
 V_s \\
 Q_s
\end{array}
\right\} 
= \sum_n\epsilon^n
\left\{
\begin{array}{c}
u_n(x)\\
g_n(x) \\
y_n(x) \\
V_n \\
Q_n
\end{array}\right\}.
\end{eqnarray}
Grouping like orders of $\e$, we find
\begin{subeqnarray}
u_0(x) & = &  \frac{h}{\mu}g_0 \left( \frac{h}{2}-y_0\right)-V_0, \slabel{u0}\\
u_1(x) & = &  \frac{h}{\mu} \left[g_1 \left( \frac{h}{2}-y_0\right) - g_0 y_1\right] + \frac{g_0^2S}{3\mu \sigma_*}\left[y_0^3 + (y_0-h)^3\right] - V_1, \slabel{u1}\\
u_2(x) & = & \frac{h}{\mu} \left[g_2 \left( \frac{h}{2}-y_0\right)-g_1y_1-g_0y_2 \right]-V_2  \slabel{u2} \\
&+&  \frac{g_0S}{3\mu \sigma_*}\left[2g_1\left[y_0^3 + (y_0-h)^3\right]+3g_0y_1\left[y_0^2 + (y_0-h)^2\right]\right],  \nonumber\\
Q_0 & = &  \frac{h^2}{2\mu} g_0 \left(\frac{h}{3}-y_0\right)   -hV_0, \slabel{Q0}\\
Q_1 & = & \frac{h^2}{2\mu}\left[g_1\left(\frac{h}{3}-y_0\right)  -g_0 y_1\right] - hV_1  -{\frac{g_0^2 S}{12\mu \sigma_*} }\left[ y_0^4 + (h-y_0)^4 - 4h  y_0^3  \right], \slabel{Q1}\\
Q_2 & = & \frac{h^2}{2\mu}\left[g_2\left(\frac{h}{3}-y_0\right)-g_1y_1-g_0y_2\right]-hV_2,  \slabel{Q2}\\
& & -{\frac{g_0S}{6\mu \sigma_*} }\left[g_1\left[ y_0^4 + (h-y_0)^4 - 4h  y_0^3\right]+2g_0y_1\left[y_0^3 + (y_0-h)^3 - 3h  y_0^2\right]\right], \nonumber \\
{\cal E}_0 & = & \frac{1}{3\mu}\int_0^\lambda g_0^2 [y_0^3 + (h-y_0)^3]\,\d x,\\
{\cal E}_1 & = & \frac{1}{3\mu}\int_0^\lambda [3y_1g_0^2[y_0^2 - (h-y_0)^2] + 2 g_0g_1[(h-y_0)^3 + y_0^3]]\, \d x \\ &-& \frac{1}{4\mu\sigma_*}\int_0^\lambda g_0^3 S[y_0^4 + (h-y_0)^4]\,\d x, \nonumber
\end{subeqnarray}
which can be solved order-by-order.  As we have shown in main body of the paper, it it not necessary to calculate ${\cal E}_2$ to determine the effect of the sign of $\e$ on the mechanical rate of work.

\section{Analytic Expansion}
\label{analytic}
The model presented in the paper is quadratic in the stress-strain rate relationship. We present here an alternate model, where the relationship is cubic,  the lowest-order model where this rheological relationship is analytic, and is written as
\begin{equation}
\dg = \frac{\sigma}{\mu} \left(1-\e \frac{\sigma^2}{\sigma_*^2} \right) \cdot
\label{shear2}
\end{equation}
We use the same notation as in the main part of the paper. The constraints on the motion, as given by  Eq.~\eqref{constraints}, remain valid here. The leading order solution is the same as in the main part of the paper, and we find 
\begin{eqnarray}
V_0 & = & 0,\\
Q_0 & = & 0,\\
{\cal E}_0& = &  \frac{h \lambda }{4\mu}\langle \sb^2 \rangle,\\
g_0 & = & \frac{3\bar\sigma }{2h}.
\end{eqnarray}
At next order, we obtain crawling with
\begin{eqnarray}
V_1 & = & -\frac{5}{32}\frac{h}{\mu \sigma_*^2} \langle \sb^3 \rangle,\\
Q_1 & = & \frac{23}{160}\frac{h^2}{\mu \sigma_*^2} \langle \sb^3 \rangle,\\
{\cal E}_1& = & -\frac{11}{80} \frac{\lambda h}{\mu \sigma_*^2}\langle \sb^4 \rangle,\\
g_1 & = & \frac{3}{80}\frac{1}{h\sigma_*^2}\left[ \langle \sb^3 \rangle-\sb^3 \right].
\end{eqnarray}
The second order solution is then given by
\begin{eqnarray}
V_2 & = & \frac{9}{1280}\frac{h}{\mu\sigma_*^4}[\langle \sb^3 \rangle\langle \sb^2 \rangle-\langle \sb^5 \rangle ],\\
Q_2 & = & \frac{21}{6400} \frac{h^2}{\mu\sigma_*^4} [\langle \sb^5 \rangle - \langle \sb^3 \rangle\langle \sb^2 \rangle],
\end{eqnarray}
and therefore we have 
\begin{equation}
\frac{\Delta Q}{Q}= \frac{21\epsilon}{920 \sigma_*^2} \frac{\langle \sb^5 \rangle - \langle \sb^3 \rangle\langle \sb^2 \rangle}{\langle \sb^3 \rangle} \cdot
\end{equation}
Now, using the same arguments and notation as in \S\ref{punchline}, we find
\begin{equation}
\langle \sb^2 \rangle =\sigma_-^2\int_0^{\alpha\lambda}g^2(u)\,\d u + \sigma_+^2 \int_{\alpha\lambda}^\lambda f^2(u)\, \d u
= \frac{(1-\alpha)}{\alpha} \sigma_+^2 I_2,
\end{equation}
\begin{equation}
\langle \sb^5 \rangle =-\sigma_-^5\int_0^{\alpha\lambda}g^5(u)\,\d u + \sigma_+^5 \int_{\alpha\lambda}^\lambda f^5(u)\, \d u
= \frac{(2\alpha^2-2\alpha+1)(2\alpha-1)(1-\alpha)}{\alpha^4} \sigma_+^5 I_5,
\end{equation}
and therefore
\begin{equation}
\frac{\Delta Q}{Q} = \epsilon\frac{21}{920}\left(\frac{\sigma_+^2}{ \sigma_*^2}\right)\frac{I_5}{I_3} \left(\frac{2\alpha^2-2\alpha+1 + (\alpha^2-\alpha)\zeta}{\alpha^2}\right),
\label{newDQ}
\end{equation}
where $\zeta=I_2I_3/I_5$. It is straightforward to show that the quadratic numerator in Eq.~\eqref{newDQ} is positive if and only $\zeta <2 $, that is, $I_2I_3< 2 I_5$. This is always true and can be shown in the same fashion as the demonstration that $I_1I_2< 2 I_3$
in \S\ref{punchline}.

In conclusion, we show that with this model as well, the sign of $\Delta Q / Q$ is the same as the sign of $\epsilon$, and therefore the mucus should be chosen to be shear-thinning ($\epsilon < 0$) in order to minimize the amount of mucus production, and therefore the overall cost of crawling. Note again that since ${\cal E}_1 <0$, this  conclusion is opposite to the one we would have obtained if mechanical work was our measure of locomotion cost.

\bibliographystyle{unsrt}
\bibliography{snail}
\end{document}